# Universal Knowledge Discovery from Big Data:

# Towards a Paradigm Shift from "Knowledge Discovery" to "Wisdom Discovery"[*]

Bin Shen

Ningbo Institute of Technology, Zhejiang University, Ningbo, China

E-mail:   tsingbin@zju.edu.cn

---

[*] "*Universal knowledge*" also can be named as "*pervasive knowledge / information*" or "*ubiquitous knowledge / information*".

# Universal Knowledge Discovery from Big Data: Towards a Paradigm Shift from "Knowledge Discovery" to "Wisdom Discovery"

*Bin Shen**

*Ningbo Institute of Technology, Zhejiang University, Ningbo, China*

*E-mail:   tsingbin@zju.edu.cn

**Abstract:** To overcome the shortcomings of knowledge discovered by traditional data mining methods, I argue that we should follow the paradigm shift from "traditional data mining" to "wisdom mining" in the era of big data, and propose Universal Knowledge Discovery from big data (UKD) as a pathway towards "wisdom" discovery. The new concept focuses on discovering universal knowledge, which exists in the statistical analyses of big data and provides valuable insights into big data. Universal knowledge comes in different forms, e.g., universal patterns, rules, correlations, models and mechanisms. To accelerate big data assisted universal knowledge discovery, a unified research paradigm should be built based on techniques and paradigms from related research domains, especially big data mining and complex systems science. Therefore, I propose an iBEST@SEE methodology. This study lays a solid foundation for the future development of universal knowledge discovery, and offers a pathway to the discovery of "treasure-trove" hidden in big data.

## Introduction

In the era of big data, data mining (DM), also called knowledge discovery in databases (KDD), has become a key means of knowledge acquisition from data. Following the classical knowledge discovery process [1] (which sequentially involves the steps of data selection, preprocessing, subsampling, transformations, pattern discovery, post-processing and knowledge utilization), data mining technology has achieved great success in the past ten years [2] in many areas, such as business, military, bioinformatics, medicine and education. Using advanced data mining methods and algorithms (e.g., pattern mining, clustering and classification methods), many previously unknown but interesting knowledge is discovered from various types of data (e.g., texts, web data, graphs and data streams) [1-5].

However, the following problems arise when applying knowledge obtained by the traditional data mining methods. Firstly, in most cases, the amount of generated knowledge is too large, so that we often cannot effectively select out suitable knowledge in practice. Taking association rule mining as an example, it is easy to generate hundreds or thousands of rules. However, it is difficult for users to comprehend or digest so many rules. Secondly, the knowledge discovered by traditional data mining method is temporary and dynamic in nature, and its validity always changes over time. Due to the lack of testing on the universality, the reliability of discovered knowledge is questioned by users. For example, suppose we obtained an association rule [3] "buy 'beer' => buy 'diaper' [$s = 2.0\%, c = 80\%$]" based on transaction data of a supermarket over a past period of time, where $s$, $c$ are the support and the confidence respectively. Since this rule was generated from historical "small" data, we do not know whether the rule still holds true in the future and under what conditions. We also do not know how the support and the confidence of the rule fluctuate over time [4], and whether this rule is applicable to other supermarkets. Thirdly, the discovered knowledge cannot sufficiently support meaningful decision-making actions, and there is a significant gap between mining results and real-world application requirements [5].

To overcome these limitations, I advocate that in the era of big data the next generation of data mining technology should focus more on mining universal knowledge, which exists in the statistical analyses of big data and provides valuable insights into big data. Generally speaking, in the history of our information age, humans have experienced two stages of data based information acquisition: the first one is using data collection and retrieval techniques to find desired information, and the second is using data mining technology to produce dynamic knowledge. I argue that the next stage will be "wisdom discovery", which adopts universal knowledge discovery technology to mine global and valuable universal knowledge (which can be easily transformed into "wisdom" in the sense of human cognition and decision making) from big data.

So, towards the vision of "wisdom discovery", a new concept called universal knowledge discovery from big data (UKD) is proposed. In order to build a solid theory foundation for the future development of UKD, the concept of UKD is defined, and various categories of universal knowledge are described. To accelerate the development of UKD, a unified research paradigm should be built based on techniques and paradigms from related research domains, which include not only data mining, but also machine discovery [6-11], big data analytics [12-25] and complex systems science [26-36]. To meet this need, a novel research paradigm called iBEST@SEE methodology is proposed. In this methodology, three types of discovery processes are also proposed: i) big-data-cycle-driven mining, ii) mechanism-cycle-driven mining and iii) combined-dual-cycle-driven mining.

Based on the new concept (i.e., UKD) and its methodology, this paper also addresses how to gain insights into big-data, which is a fundamental issue and urgent problem for all scientific disciplines [37-39]. Overall, this work serves three purposes. First, UKD is highlighted as a pathway to the discovery of "treasure-trove" hidden in big data. Second, I advocate that an interdisciplinary and unified research paradigm for UKD should be built. Therefore, an iBEST@SEE methodology is proposed. Last, this research lays a solid foundation for the future development of UKD technology, and aims at underpinning a new wave of universal knowledge discovery in big data.

## Technical origins

To integrate methods in related disciplines to build advanced UKD technology, we need to sort out paradigms and techniques adopted in related fields. Therefore, in this section, I review technical origins of universal knowledge mining, which include machine discovery, big data analytics, complex systems science, etc.

**From machine discovery to data mining, and then big data mining**

As early as the 1970s, Langley [6, 7] had developed six versions of the BACON system one after another, which are able to automatically rediscover some physical and chemical laws using heuristic searching methods. Soon afterwards, based on inductive learning and logical reasoning, a number of systems, such as STAHL [8], FAHRENHEIT [9] and IDS [10], were developed in this field. In China, Chinese scholars, such as Wu [11], have made fruitful achievements in mechanical theorem proving in geometries. Thus, in the 1980s, the research direction of machine discovery found focus. However, these studies have some limitations. Firstly, the algorithms can only deal with pure and small data as the input and need to be accompanied by a certain searching direction; otherwise the algorithms are infeasible due to huge searching space. Secondly, the systems only rediscovered some laws already known, and few important and new scientific laws have been discovered by such systems.

Since the mid-1990s, data mining research has avoided the limitations of machine discovery by discovering meaningful patterns from the statistical perspective instead of scientific laws. Thus, a large number of effective data mining algorithms have been developed, and data mining which integrates related disciplines (e.g., statistics and machine learning) has achieved great success in practice.

The advent of the era of big data presents a new opportunity for the future development of data mining technology. Currently, research in big data mining (or big data analytics) commonly focuses on developing advanced technology to deal with the technical aspects of big data, e.g., large volume [12, 13], high velocity, and /or high variety [14]. These studies mainly are concerned with the performance, the scalability and the scope of proposed methods, but neglect fundamental changes in thinking brought about by the shift from processing small data to tackling big data [15]. Most data mining researchers have not been aware of the potential great changes in research paradigm and mining methods as they shift from "knowledge discovery from small data" to "universal knowledge discovery from big data".

For instance, McKinsey Global Institute pointed out that big data is the next frontier for innovation, competition and productivity [16]. However, the methods listed in the report are still traditional ones. Cohen et al. [17] highlighted Magnetic, Agile and Deep (MAD) skills for big data analysis, which depart from traditional enterprise data warehouses. Zikopoulos et al. [18] introduced popular Hadoop distributed processing platform for big data analytics. All three articles [19, 20] presented at the session of "Big data frameworks" of SIGKDD 2013 focus on high performance processing of big data. In the survey paper of big data mining [21], Fan et al. mainly discussed the computability in big data processing and how to process various types of big data.

Although the treatability of big data is an important aspect of big data management and mining, it is not enough to discover "treasure-troves" hidden in big data. Obviously, compared with "small" data mining, big data mining doesn't merely mean that the amount of data that can be processed increases, but also, more importantly, means a general shift in thinking (e.g., focusing on data externality and highlighting correlations [15]) as well as a new data mining research paradigm. In this way, we can not only process big data as easily as small data, but also gain insights cleverly based on externalities of big data [22]. I consider that the following works are representative of big data analytics.

For instance, during the 2014 Chinese Spring Festival, based on big data of more than 200 million smartphones, China Central Television exhibited the macroscopic statistical regularities of population movements within China during this period. Based on such knowledge, related public agencies are able to efficiently schedule resources of passenger transportation, and individuals can better arrange their travel plans [23]. This work takes full advantage of the externalities of smartphone trajectory data. In addition, Zhu et al. [24] analyzed the big data of Chinese food recipes, and found that the geographical proximity rather than climate similarity is a crucial factor determining the similarity of regional cuisines. Here, the data analyzed is the entire dataset of Chinese recipes, rather than sampling data. Furthermore, if big data of recipes from all countries around the world were analyzed, the conclusion would be more persuasive. Zhang et al. [25] used trajectories from a fleet of GPS-equipped taxicabs to detect gas station visits and measure the time spent in real time, and then estimate overall gas usage of the city. In this case, the system has the capability of real-time scheduling and decision-making based on the comprehensive dynamic perception data of an entire smart city.

The above practical applications of big data are quite good. However, overall, applications of big data analytics are still in the stage of trial and error. An ideal application paradigm of big data has not been formed yet.

**Scientific progress and research paradigm of big data based complex systems science**

In China, early in the 1980s, local pioneer scientists Qian [26, 27] and Xu [28] et al. began to advocate the necessity of system science and proposed some important thoughts and ideas. However, without the relevant data, it was difficult to prove these ideas at that time. With the rise of the big data technologies in recent years, especially the techniques of big data collection, storage and analysis, it has been feasible to explore hidden mechanisms and laws behind big data and verify some thoughts based on large datasets, which has led to a new surge of complex systems research.

For example, the well-known small world model [29] and the scale-free network theory [45] are both based on large-scale empirical datasets, e.g., the actor cooperation network and the World Wide Web network datasets. Besides, the research on the inverse problem of complex network (i.e. uncovering the topology of unknown networks according to observation data which reflects network dynamics), such as link prediction [30], are all required to be verified by big data of large scale networks. For another example, by analyzing the flow trajectories of 464,670 dollar bills, Brockmann [31] uncovered the knowledge of dynamic and statistical properties of human travel on a large scale, and discovered the dispersal of bank notes matches a continuous time random walk process incorporating scale free jumps as well as long waiting times between displacements. Likewise, Gonzalez [32], Yan [33] and Peng [34] reached some valuable conclusions by analysing the trajectories of over 100 thousand mobile users, 230 volunteers' 6-week travel diaries and the movement paths of 15.8 thousand Shanghai taxies respectively.

Based on the above cases, we may conclude that complex systems science adopts a different technical paradigm, compared with the data mining research field. Its characteristics are summarized below [35].

1) A systems theory perspective

System science has abandoned reductionism and embraces holism. Complex network, as a rapidly growing field of system science, offers a fresh perspective on the use of networks for system modeling and analytics. On this basis, core issues, such as system dynamics and function emergence, are being researched. Macroscopic phenomena can be reproduced via setting appropriate micro-level simple rules for system dynamics.

2) The viewpoints of evolution and dynamics

System science takes the perspective of evolution and dynamics on systems. It not only discusses the dynamics of system topology, but also cares about various dynamic phenomena on systems. Taking complex network research as an example, it explores a variety of dynamic mechanisms based on certain network topologies, such as the spread of infectious diseases, cascading failures, network congestion and crowd-powered socially embedded search engine.

3) Extensive using statistical physics methods

Statistical physics has played an important role in complex systems research and provided a variety of effective approaches, e.g., the mean-field theory, the percolation theory, the master equation and the Fokker-Planck equation. At the same time, the complexity research also concerns physical processes taking place in systems and the emergence of physical properties of systems, such as Bose-Einstein condensation and random walks on complex networks.

4) Combination of simulations and statistics on big data

System science tries to find and validate statistical laws based on big data, and discover the internal mechanisms of systems. Computer simulation is also a frequently used tool in the research of system science.

In summary, the research paradigm of complex systems science has apparent advantages in system modeling, dynamics analysis, mechanism exploring, etc. However, there is a bottleneck in big data processing and analytics for system science. Wang et al. [36] considered that systematically using empirical data to explore systems is still in its preliminary stage, and the key issue is the lack of effective theoretical framework for well-targeted data accumulation and analytics. But in fact, there are many advanced methods and valuable experiences in the data mining field for data acquisition, modeling and processing [35]. Therefore, it is necessary to integrate the methods in complex systems science and data mining for solving big-data-assisted discovery problems. This issue will be discussed in the section on methodology in detail.

## Materials

**What are universal knowledge and universal knowledge discovery from big data**

*Universal KNowledge* (UKN)[1] in big data is defined as knowledge and laws with a certain degree of universality and immutability in the statistical analyses of big data. In this definition, *universality* implies that the phenomenon highlighted by the knowledge exists extensively in nature and society, and *immutability* implies that the phenomenon remains unchanged under certain conditions with the evolution of the universe. With its repeatability and predictability, UKN provides insights into big data. Based on the definition of UKN, the concept of *universal knowledge discovery from big data (UKD)* is quite straightforward. It is defined as the process of discovering universal knowledge from big data.

---
[1] Universal knowledge also can be named as "pervasive knowledge " or "ubiquitous knowledge".

To understand the concept of UKN and its applications easily, in the following paragraphs, I give three examples to illustrate the current progress towards this direction.

Example 1. Universal knowledge about viral marketing

A marketing manager wants to set a number of individual nodes (i.e., monitoring staff) in a social network to monitor the status of a viral marketing campaign, and forecast whether or not a phenomenon designed by the campaign will "go viral". Chesney [40] discovered some UKN in multiple evolutionary simulations. For example, the number of individual nodes adopting a viral behavior follows an s-curve, and whether the adoption proportion exceeds a critical value or not will determine its prevalence. In addition, a node's predictive accuracy is negatively related to its geodesic distance to the nearest member of the critical mass (a small group of early adopters), and positively related to its centrality. Based on these findings, marketers can take appropriate actions, e.g., selecting individuals with high network centrality which are scattered in different clusters as observation nodes.

Example 2. Universal metaphors and its applications

After analyzing web based big data in various languages, we may find many frequent metaphors, such as the comparison of "father" to "sun". Because such metaphors are frequently used in various languages, they can be called universal metaphors. According to Cancho's work [41], a "small world" network of human language can be built, if any two frequent co-occurring words, which include pairs of tenors and vehicles of universal metaphors, are connected with each other. We also know that human brains use networks of neurons connected by synapses to store the "small world" of human language [42]. Therefore, we can speculate that the neural connections of concepts "father" and "sun" are universal in human brains. Exploring the origin of such universal neural connections is helpful for the in-depth study of human languages and brain science.

Example 3. Universal small world phenomenon and power law distribution

Milgram [43] found the phenomenon of small world from his famous small-world experiment, and Watts [44] discovered the same phenomenon from empirical data of actor collaboration networks, power grids, etc. Now, the existence of such phenomenon has been confirmed in more and more fields, such as Internet and the spread of diseases. Barabási [45] found that the distribution of node degree of a scale-free network follows a power law. We now know that such power law distribution exists extensively in nature and society. Examples include the Zip's-law-like word frequency distribution, the Pareto distribution of social wealth, the Gutenberg-Richter power law distribution of earthquake sizes, and the intensity distribution of solar flares. It shows that this law is in line with the evolution of the universe.

From the above examples, we can find that the discovery of such UKN is meaningful, and exploring the origins of universal phenomenon highlighted by the UKN leads to profound insights.

[Figure 1]

It is necessary to point out that universal knowledge includes but is not limited to scientific laws (as shown in Figure 1). That is to say, scientific laws can be regarded as a special kind of universal knowledge. Compared with scientific laws, universal knowledge stands in the statistical analyses of big data and encompasses a much broader scope. Universal knowledge is also very meaningful. For example, using large-scale mobile phone data, Gao et al. [46] suggests that communication traffic surges after disasters and the dominant component of this traffic is between eye-witnesses and their significant friends and relatives. The discovery of such universal knowledge is quite useful for emergency management.

**Categories of universal knowledge**

Universal knowledge can be classified into the following non-exhaustive representative categories, which can be extended in accordance with human cognition expansion in the future.

- **Universal knowledge in data distributions**

The distribution of big data helps us to characterize big data. Common data distributions include normal / Gaussian distribution, log-normal distribution, exponential distribution, Poisson distribution, power law, power-law with exponential cutoff, shifted power-law, stretched exponential distribution, Weibull distribution, fat-tailed distribution, and so on. For example, Lévy flights are expected in places where prey is scarce, whereas a Brownian strategy is more likely to be adopted where prey is abundant [47].

- **Universal knowledge towards characteristics of systems**

Common characteristics of systems include the following: self-similarity and fractal nature, small world, scale free, bursts and so on. For example, Barabási [48] shows that the pattern of bursts appears in a wide range of human daily activities, from web browsing to letter writing, from Wikipedia editing to online trading. For another example, financial markets normally have long-term memory.

- **Universal knowledge about indexes**

People may find that there are certain correlations between phenomena and their indicators. These correlations include but are not limited to causal relationships, as discussed in the part of universal correlations below. In order to indicate the fluctuation of a certain phenomenon, we may combine correlated indicators to build indexes. For example, China's "Li Keqiang index" combines electricity consumption, railway freight and lending to measure the country's economic growth.

- **Universal patterns**

Many kinds of patterns have been developed in the data mining field, such as frequent co-occurrence patterns and behavior patterns [49]. If these patterns exist in the big data of a system, we can call them universal patterns. For further example, since words "I" and "am" frequently appear in sequence in English, "I am" can be called a universal sequential pattern. For another example, because the comparison of "father" to "sun" is frequently used in different languages, it can be seen as a universal metaphor.

- **Universal rules**

Based on big data, we can obtain universal rules, e.g., universal association rules, universal IF-THEN rules and universal intervention rules. For example, by analyzing the birth defects monitoring data of a certain region (e.g., region "A"), we find that the proportion of birth defects of newborns decreases obviously if the intervention of early pregnancy folic acid supplements is adopted. Thus, an intervention rule is obtained as below [50].

"Folic acid (deficiency → sufficiency) => Birth defects (Yes → No), support=0.33".

Then, if the validity of this rule is confirmed by testing in different regions around the world, this rule can be called a universal intervention rule.

- **Universal correlations**

Universal correlations exist in the statistical sense of big data. Universal correlations are not restricted to universal causality, but also include much broader universal positive / negative correlations and universal co-occurrence relationships, etc. Two phenomena, which are positive / negative correlated or co-occurrence, may be caused by the same reason or merely co-occur. However, finding such correlations is also meaningful. For instance, based on the relevance of "flu" related online search queries and the breakout of seasonal pandemic influenza, Google Flu Trends can predict the occurrence of influenza in a more timely and accurate fashion [51].

- **Universal models**

If we use machine learning models to learn universal phenomena inductively, or adopt mathematical formulas or physical models, etc. to characterize universal phenomena, universal machine learning models, universal mathematical models, universal physical models, etc. can be obtained correspondingly. For instance, based on the unsupervised deep learning method [52], Google's artificial brain learns to identify a cat. This artificial neural network is universal, because it can recognize any cat.

- **Universal mechanisms**

Discovering universal mechanisms, which are behind universal phenomena in big data, has significant implications for gaining insights. For example, Barabási [45] explained that growth and preferential attachment are mechanisms common to a number of self-organizing complex networks, and they make vertex connectivities in large networks following power-law distributions at the system level.

**Comparison of universal knowledge and knowledge generated by traditional data mining methods**

Based on the above discussion about various categories of universal knowledge, we make a comparison between UKN and knowledge discovered by the traditional data mining techniques (traditional knowledge for short) as shown in Table 1. From the table, we can find that UKN has significantly different characteristics compared with traditional knowledge. 1) The scope is broader and its categories are more diverse. 2) UKN is extracted from the perspective of the whole system and is normally described at the macro level. 3) Both precise and general descriptions are acceptable. 4) UKN is mined from and tested in big data. 5) UKN affords easy comprehension and insights, and has a strong actionability.

## Methodology

**Comparison of research paradigms of big data mining and system science**

In the era of big data, facing a novel class of data mining tasks, i.e., universal knowledge mining, it is urgent to do an interdisciplinary integration of research paradigms and techniques of related fields. For this purpose, in Table 2, a comparison is made between big data mining and system science, which are two main technical origins of UKD. The comparison shows that the approaches in these two fields have high mutual complementarity when treating big data. That is, complex systems science research is good at discovering macroscopic laws and internal mechanisms from big data; by contrast, data-driven big data mining specializes in handling various types of big data. So it is necessary to integrate the related fields, especial big data mining and complex systems science, to build a unified and interdisciplinary research paradigm for big-data-assisted universal knowledge discovery.

**iBEST@SEE methodology: a unified and interdisciplinary paradigm for UKD**

To meet the above need of integrating related fields, in this subsection, I propose iBEST@SEE methodology and its discovery processes, which provide a unified and interdisciplinary research paradigm for universal knowledge discovery.

Firstly, we use "big data mining" (containing the techniques of data-driven analytics and statistics) and "system science" (including complexity thinking, system modelling and dynamics, etc.) to generalize the paradigms of these two fields respectively. In addition, we need to adopt the method of "simulation" in system science, and draw on some classical methods (or paradigms) in other disciplines or fields, as described below.

1) Simulation

This means the use of computer algorithms or other simulation systems to simulate real systems. This approach is especially fit for multi-agent simulation, system evolution simulation, numerical simulation, etc., can get simulation results quickly and has the advantage of low cost.

2) Controlled experiment

This means that an observer tests hypotheses by looking for changes brought on by manipulating one or more independent variables while all other variables are held constant [53]. It includes the steps of putting forward premises (including hypotheses), experimental operation, result verification, etc.

3) Theoretical Analysis

This means the use of field theories accompanied by mathematical, physical and other tools in modelling, theoretical analysis, deduction, proof, etc.

4) Intelligence

Intelligence comes from machine intelligence (which comes from expert systems, machine learning, web intelligence, etc.), human intelligence (which can be individual and also collective intelligence), intelligence generated from human-computer interaction, etc.

5) Empirical Research

Big data based UKD is a typical empirical problem. Therefore, we need to use an empirical cycle [54] for hypothesis testing, which includes observation (i.e., collecting and organizing empirical facts), induction (i.e., formulating hypothesis), modelling, deduction (i.e., deducting consequences of hypothesis as testable predictions), testing (i.e., testing the hypothesis with new empirical material) and evaluation. In each cycle, hypotheses and models are adjusted gradually to better fit the empirical facts.

On the basis of the above discussion, I propose a novel research paradigm for UKD called iBEST@SEE methodology. "iBEST@SEE" stands for " 'intelligence' + 'Big Data Mining' + 'Experiment' + 'Simulation' + 'Theory Analysis' @ 'System Science' + 'Extended Empirical Research Framework' ", where new phrase "iBEST@SEE" is formed by the underlined letters. Here, "@" emphasizes that the methodology systematically organizes various methods before "@" by adopting the extended empirical research framework from the system science perspective. The lowercase letter "i", distinguished from the other uppercase letters, is used to highlight the core role of human intelligence in this creative discovery process.

The traditional knowledge discovery process [1] is unable to fully satisfy the requirements of UKD. Therefore, in iBEST@SEE methodology, three types of discovery processes are proposed for UKD as below.

**[Figure 2]**

I. Big-data-cycle-driven mining

Big-data-cycle based mining may be applied directly, under the condition of appropriate datasets and certain mining target of universal knowledge. It consists of four steps: 1) system observation and scheme design, 2) big-data-cycle-driven analytics, 3) UKN understanding, evaluation and management, and 4) action adoption, as shown in Figure 2A.

In the first step, "system observation and scheme design" observes emergent macroscopic phenomena and functions of the system, and arranges the whole mining process, which includes setting the mining target (i.e., a certain type of UKN), using big data and complexity thinking to design a scheme for UKD, etc. The second step "big-data-cycle-driven analytics" is the kernel of the whole process, as shown in the left cycle of Figure 3. The big data cycle consists of five sub-steps: ①big data accumulation and acquisition, ②big data preprocessing, ③big-data-driven analytics, ④universal phenomenon discovery, and ⑤forecasting, verification and evaluation. These five sub-steps constitute a spiraling process, in order to discover universal phenomena from big data. The third step "UKN understanding, evaluation and management" is the basis of knowledge utilization. It assesses discovered UKN, which includes its universality, usability, etc., and treats it as a strategic asset. The final step "action adoption" generates strategies based on discovered UKN and domain knowledge, and then takes actions.

The process of big-data-cycle-driven mining is suitable for searching, mining and verifying UKN with a clear target, such as verifying whether the data follows a certain distribution, mining frequent universal patterns from the data set, building a universal neural network model for the given data set, etc.

II. Mechanism-cycle-driven mining

If empirical big data is missing, we may adopt "mechanism-cycle-driven mining" to explore experimental data generated by experiments and simulations. It contains the following five steps as shown in Figure 2B: 1) system observation and scheme design, 2) mechanism-cycle-driven analytics, 3) putting forward knowledge hypothesis and further verification, 4) UKN understanding, evaluation and management and 5) Action adoption.

Here, the first step is the same as the one in big-data-cycle-driven mining. The second step "mechanism-cycle-driven analytics" is the core of this process and is shown in the right cycle of Figure 3. This step consists of five sub-steps: ①system abstraction and modeling, ②theoretical analysis and hypothesis proposition, ③simulations and controlled experiments, ④experimental data analysis and ⑤the emergence of universal phenomenon. These five sub-steps form a spiraling cycle called mechanism cycle, which adopts the empirical research framework. If the deduction of hypotheses explain the observed experimental phenomena, the hypotheses are acceptable; otherwise, the hypotheses have to be revised gradually until the phenomena can be well-explained. At the same time, the emergent phenomena in the experiment should match the observed reality. Thus, hypotheses and related knowledge (which are called knowledge hypotheses) can be put forward for further verification in big data in the third step. If knowledge hypotheses are acceptable in big data, they become valuable universal knowledge and can be fully utilized in the next two steps, which are the same to the ones in big-data-cycle-driven mining.

This process is designed to explore various classifications of UKN, especially internal mechanisms, mathematical and physical models, etc.

III. Combined-dual-cycle-driven mining

The process of combined-dual-cycle-driven mining (Figure 2C) is similar to that of big-data-cycle-driven mining (Figure 2A). The difference is that the former adopts "combined-dual-cycle-driven analytics", which combines both mechanisms in two cycles as shown in Figure 3, instead of one cycle driven analytics.

**[Figure 3]**

It can progressively discover both statistical universal phenomena and in-depth internal mechanisms in big data. Here, the collaboration of two mechanisms of two cycles works with "two adaptations" and "two matches". "Two adaptations" have two aspects. (i) "System abstraction and modeling" in the mechanism cycle supervises "big data accumulation, acquisition and preprocessing". "Big data accumulation, acquisition and preprocessing" and "big-data-driven analytics" reflect "system abstraction and modeling". (ii) "Universal phenomenon discovery" in the big data cycle supports "theoretical analysis and hypothesis proposition" which is contained in the mechanism cycle. On the other hand, "theoretical analysis and hypothesis proposition" guides "universal phenomenon discovery". "Two matches" suggest that (i) universal phenomena discovered in the big data cycle should match to universal phenomena appeared in the simulations and experiments, and (ii) the experimental phenomena should be consistent with the phenomena observed in the reality. If "two matches" exist, the proposed hypotheses are acceptable; otherwise, amend them until "two matches" are proved.

## Discussions and conclusions

In the era of big data, the flood of data in various fields has been and is still being generated constantly. In result, the deluge of big data has attracted increasing attention in recent years [55-58]. Data-intensive scientific discovery has become the fourth paradigm for scientific exploration, after experimental research, theoretical research and computer simulations, according to the pioneer scientist Jim Gray [37]. Big data is regarded as a big deal since it promises to change the world [38]. However, how to deal with big data to provide big insights and make a big impact is still an open question [39].

As mentioned earlier, machine discovery aims at discovering scientific laws from data automatically. Recent research has advanced the automation of the hypothetico-deductive method and the recording of experiments. However, heuristic-searching-and-inductive-learning based machine discovery is only capable to process clean and smaller data due to large search spaces [6-11]. Data mining has achieved great success in practice by integrating related disciplines (e.g., statistics and machine learning) [1-5]. However, without testing on the universality, the knowledge discovered by traditional DM methods is temporary and dynamic in nature. In big data mining, computer scientists commonly focus on technical aspects (e.g., the performance and the scope) of the proposed methods [16-21], but neglect fundamental changes in thinking as they shift from processing small data to tackling big data, such as focusing on data externality and highlighting correlations [15]. Currently, there have been several successful attempts of gaining insights from big data in complex systems science [44-48], but, overall, big-data-assisted scientific discovery is still in its earliest exploratory stage. Most importantly, no unified research paradigm has been established yet for big data discovery.

To meet the above challenges, UKD is proposed. The new concept focuses on discovering universal knowledge, which exists in the statistical analyses of big data and provides valuable insights into big data. Universal knowledge has significantly different characteristics compared with knowledge discovered by traditional data mining techniques. To accelerate universal knowledge discoveries from big data, it is urgent to integrate the related fields, especially big data mining and complex systems science, to build a unified and interdisciplinary research paradigm for developing advanced UKD technology. To meet this need, I propose a novel research paradigm for UKD called iBEST@SEE methodology. In iBEST@SEE methodology, I propose three types of discovery processes: i) big-data-cycle-driven mining, ii) mechanism-cycle-driven mining and iii) combined-dual-cycle-driven mining.

The primary goal of this research is to underpin a new wave of universal knowledge discovery in big data. The work proposed here builds a foundation for the future development of UKD technology. I envision that the universal knowledge discovery paradigm will become a key basis to address the spectrum of big data challenges. In the light of UKD, it is foreseeable that more and more universal knowledge with a big impact will be discovered from big data in the future.

## References

1. Fayyad U, Piatetsky-Shapiro G, Smyth P (1996) From data mining to knowledge discovery in databases. AI Magazine 17: 37-54.
2. Han J, Kamber M, Pei J (2011) Data Mining: Concepts and Techniques. Burlington, MA: Morgan Kaufmann, ed. 3.
3. Agrawal R, Imielinski T, Swami A (1993) Mining association rules between sets of items in large databases. Proceedings of the ACM SIGMOD conference on management of data 1993 207-216.
4. Shen B, Yao M, Wu Z, Gao Y (2010) Mining Dynamic Association Rules with Comments. Knowledge and Information Systems 23: 73-98.
5. Cao L (2012) Actionable knowledge discovery and delivery. Wiley Interdisciplinary Reviews: Data Mining and Knowledge Discovery 2: 149-163.
6. Langley P (1978) BACON.1: A general discovery system. Proceedings of the Second Biennial Conference of the Canadian Society for Computational Studies of Intelligence 173-180.
7. Langley P (1979) Rediscovering physics with BACON.3. Proceedings of the Sixth International Joint Conference on Artificial Intelligence 505-507.
8. Zytkow JM, Simon HA (1986) A theory of historical discovery: the construction of componential model. Machine Learning 1: 107-136.
9. Zytkow JM, Zhu J, Hussam A (1990) Automated discovery in a chemistry laboratory. Proceedings of AAAI'90 889-894.
10. Nordhausen B, Langley P (1993) An integrated framework for empirical discovery. Machine Learning 12: 17-47.
11. Wu WS, Jin XF, Wang DM (1994) Mechanical theorem proving in geometries: basic principles. Berlin: Springer-Verlag.
12. Lin J, Ryaboy D (2012) Scaling big data mining infrastructure: the twitter experience. SIGKDD Explorations 14: 6-19.
13. Rajaraman A, Ullman JD (2012) Mining of massive datasets. Cambridge: Cambridge University Press.
14. IBM (2012). What is big data? Available: http://www-01.ibm.com/software/data/bigdata/. Accessed 2 December 2012.

15. Mayer-Schönberger V, Cukier K (2013) Big Data: A Revolution That Will Transform How We Live, Work, and Think. Boston: Houghton Mifflin Harcourt.
16. McKinsey Global Institute (May 2011). Big data: The next frontier for innovation, competition, and productivity. Available: http://www.mckinsey.com/insights/business_technology/big_data_the_next_frontier_for_innovation. Accessed 20 May 2011.
17. Cohen J, Dolan B, Dunlap M, Hellerstein JM, Welton G (2009) MAD skills: new analysis practices for big data. Proceedings of VLDB'09 1481-1492.
18. Zikopoulos P, Eaton C (2011) Understanding big data: analytics for enterprise class hadoop and streaming data. New York: McGraw-Hill Osborne Media Publisher.
19. Han WS, Lee S, Kim KP, Park K, Lee JH, Kim MS, Kim J, Yu H (2013) TurboGraph: a fast parallel graph engine handling billion-scale graphs in a single PC. Proceedings of KDD'13 77-85.
20. Canny J, Zhao H (2013) Big data analytics with small footprint: squaring the cloud. Proceedings of KDD'13 95-103.
21. Fan W, Bifet A (2013) Mining big data: current status, and forecast to the future. SIGKDD Explorations 14: 1-5.
22. Zhou T (29 Jan 2013) In the big data era, China has not lagged behind. Available: http://blog.sciencenet.cn/blog-3075-657481.html. Accessed 29 Jan 2013. (in Chinese)
23. Tang C (08 Feb 2014) http://blog.sciencenet.cn/home.php?mod=space&uid=287179&do=blog&id=765603. Accessed 08 Feb 2014. (in Chinese)
24. Zhu YX, Huang JM, Zhang ZK, Zhang QM, Zhou T, Ahn YY (2013) Geography and similarity of regional cuisines in China. PLoS ONE 8: e79161.
25. Zhang F, Wilkie D, Zheng Y, Xie X (2013) Sensing the pulse of urban refueling behavior. Proceedings of UbiComp'13 13-22.
26. Qian X (1981) A rediscussion on the system of system science. Systems Engineering – Theory & Practice 1: 1-3. (in Chinese)
27. Qian X, Yu J, Dai R (1990) A new field of science: open complex giant system and its methodology. Nature magazine of China 13: 3-10. (in Chinese)
28. Xu G, Gu J, Che H (2000) System science. Shanghai: Shanghai scientific and technological education press. (in Chinese)
29. Watts DJ, Strogatz SH (1998) Collective dynamics of 'small-world' networks. Nature 393: 440-442.
30. Lü L, Zhou T (2013) Link prediction. Beijing: Higher education press. (in Chinese)
31. Brockmann D, Hufnagel L, Geisel T (2006) The scaling laws of human travel. Nature 439: 462-465.
32. González MC, Hidalgo CA, Barabási AL (2008) Understanding individual human mobility patterns. Nature 453: 779-782.

33. Yan XY, Han XP, Wang BH, Zhou T (2013) Diversity of individual mobility patterns and emergence of aggregated scaling laws. Scientific Reports 3: 2678.
34. Peng CB, Jin XG, Wong KC, Shi MX, Lio P (2012) Collective human mobility pattern from taxi trips in urban area. PLOS One 7: e34487.
35. Shen B (2014) A comparative study and an integration of research paradigms of complex networks and data mining. Complex Systems and Complexity Science 11: 48-52. (in Chinese)
36. Wang B, Zhou T, Zhou C (2012) Statistical physics research for human behaviors, complex networks and information mining. Journal of University of Shanghai for Science and Technology 34: 103-117. (in Chinese)
37. Hey T, Tansley S, Tolle K, eds (2009) The Fourth Paradigm: Data-intensive Scientific Discovery. Redmond: Microsoft Research.
38. Shaw J (March / April 2014). Why "big data" is a big deal: Information science promises to change the world, Harvard Magazine. Available: http://harvardmagazine.com/2014/03/why-big-data-is-a-big-deal. Accessed 28 April 2014.
39. Lazer D, Kennedy R, King G, Vespignani A (2014) The parable of Google Flu: Traps in big data analysis. Science 343: 1203-1205.
40. Chesney T (2014) Networked individuals predict a community wide outcome from their local information. Decision Support Systems 57: 11-21.
41. I Cancho RF, Sole RV (2001) The small world of human language. Proceedings of the Royal Society B (Biological Science) 268: 2261-2265.
42. Pulvermüller F (2003) The neuroscience of language: on brain circuits of words and serial order. Cambridge: Cambridge University Press.
43. Milgram S (1967) The small-world problem. Psychology Today 1: 61-67.
44. Watts DJ, Strogatz SH (1998) Collective dynamics of 'small-world' networks. Nature 393: 440-442.
45. Barabási AL, Albert R (1999) Emergence of scaling in random networks. Science 286: 509-512.
46. Gao L, Song C, Gao Z, Barabási AL, Bagrow JP, Wang D (2014) Quantifying information flow during emergencies. Scientific Reports 4: 1-6.
47. Viswanathan GM (2010) Fish in Lévy-flight foraging. Nature 465: 1018-1019.
48. Barabási AL (2011) Bursts: The hidden patterns behind everything we do, from your e-mail to bloody crusades. New York: Penguin Group.
49. Cao L, Yu PS (2012) Behavior computing: modeling, analysis, mining and decision. Berlin: Springer.
50. Tang C, Zhang Y, et al. (2008) A survey on mining kinetic intervention rule from sub-complex systems. Journal of Computer Application 28: 2732-2736. (in Chinese)
51. Ginsberg J, Mohebbi MH, Patel RS, Brammer L, Smolinski MS, Brilliant L (2009) Detecting influenza epidemics using search engine query data. Nature 457: 1012-1014.
52. Bengio Y (2009) Learning deep architectures for AI. Foundations and trends in machine learning 2: 1-127.

1753. (01 Jan 2014) What is a controlled experiment? Available: http://www.innovateus.net/innopedia/what-controlled-experiment. Accessed 30 June 2014.
54. De Groot AD (1969) Methodology: foundations of inference and research in the behavioral sciences. Belgium: Mouton & Co.
55. Lynch C (2008) Big data: how do your data grow? Nature 455: 28-29.
56. Marx V (2013) Biology: the big challenges of big data. Nature 498: 255-260.
57. Bell G, Hey T, Szalay A (2009) Beyond the data deluge. Science 323: 1297-1298.
58. Akil H, Martone ME, Van Essen DC (2011) Challenges and opportunities in mining neuroscience data. Science 331: 708-712.

53. (01 Jan 2014) What is a controlled experiment? Available: http://www.innovateus.net/innopedia/what-controlled-experiment. Accessed 30 June 2014.
54. De Groot AD (1969) Methodology: foundations of inference and research in the behavioral sciences. Belgium: Mouton & Co.
55. Lynch C (2008) Big data: how do your data grow? Nature 455: 28-29.
56. Marx V (2013) Biology: the big challenges of big data. Nature 498: 255-260.
57. Bell G, Hey T, Szalay A (2009) Beyond the data deluge. Science 323: 1297-1298.
58. Akil H, Martone ME, Van Essen DC (2011) Challenges and opportunities in mining neuroscience data. Science 331: 708-712.

# Figure Legends

**Figure 1**. **The difference between scientific laws and universal knowledge**

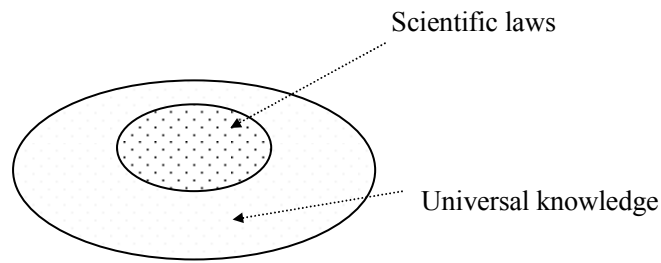

**Figure 2**. **The processes of big-data-cycle-driven UKD (A), mechanism-cycle-driven UKD (B) and combined-dual-cycle-driven UKD (C).**

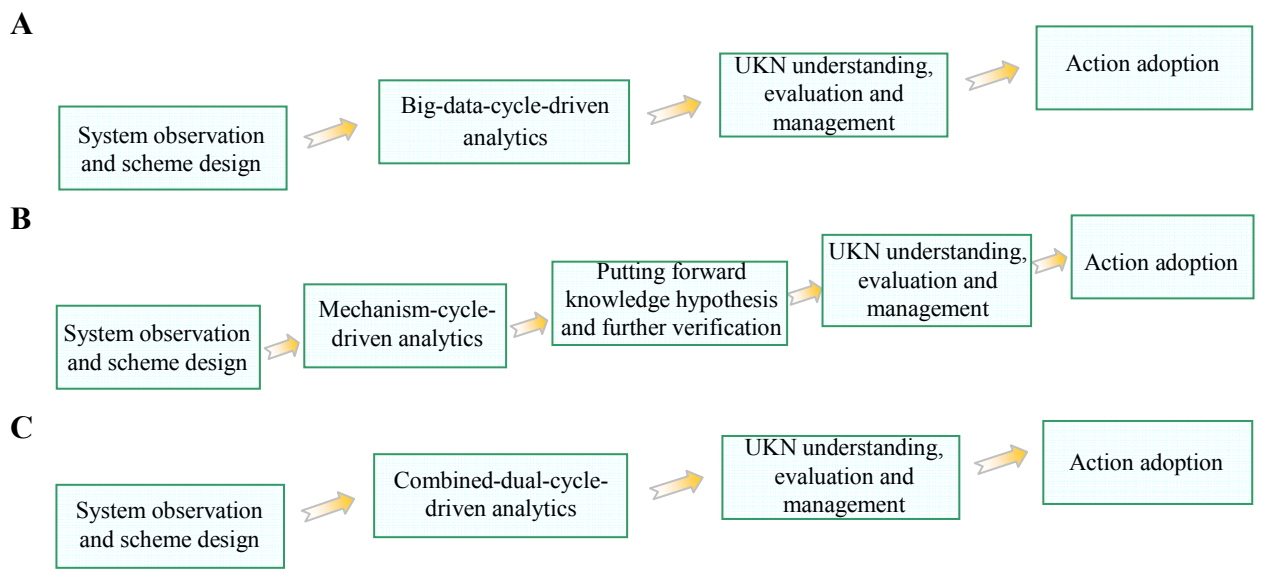

**Figure 3. Combined-dual-cycle-driven analytics.**

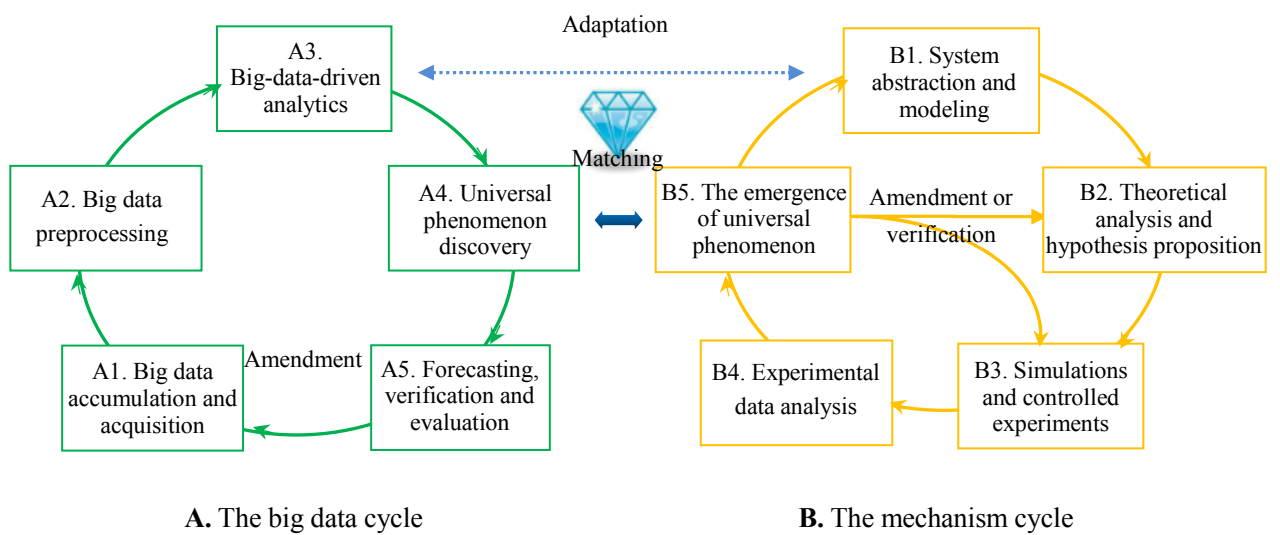

# Tables

**Table 1.** A comparison between characteristics of two types of knowledge**.**

| Aspects | Traditional knowledge | Universal knowledge |
| --- | --- | --- |
| Types | Typically patterns, rules and machine learning models | Not only includes patterns, rules and machine learning models, but also contains distributions, characteristics, indexes, correlations, mathematical models, etc. |
| Description level | Generally at the micro level or meso-level | Using a systematic perspective and describing at the macro level |
| Degree of precision | Most knowledge is precise and is described in detail | Both precise and general descriptions are acceptable |
| Data | Typically local and small data | Global big data |
| Evaluation | Some samples | Big data |
| Universality | Without further universal testing | A certain degree of universality |
| Comprehensibility | Commonly hard to understand | Affords easy comprehension and insights |
| Actionability | Typically weak | Strong |
| Calculation | Tend to be accurately calculated | Estimates are also effective |

Traditional knowledge stands for knowledge generated by traditional data mining methods.

**Table 2.** Comparison of big data mining and system science.

| Aspects | Big data mining | System science |
|---|---|---|
| Focus | Treatability of big data, performance of platforms and algorithms, interestingness of discovered knowledge, etc. | Three elements of systems, i.e., system structure, dynamics and functions |
| Perspectives | Big data based induction, statistics and machine learning | Systematic, evolutionary and network perspectives |
| Thinking | Collecting and using a complete data set, utilizing externalities of big data, effective estimates, looking for correlations, etc. | Complexity thinking, e.g., self-organization, emergence, self-similarity and chaos |
| Models | Typically machine models | Mathematical and also physical models |
| Methods | Statistical analysis, data mining and machine learning algorithms | Statistical physics, computer simulations, algorithms, etc. |
| Knowledge categories | Machine learning models, patterns and rules | Mathematical and also physical models, universal laws and internal mechanisms |
| Virtues | Specializing in handling various types of big data efficiently and discovering hidden statistical patterns | Having advantages in modeling systems, and discovering universal laws and internal mechanisms |

# 普适知识挖掘：
# 从"知识发现"到"智慧发现"的技术进路[*]

沈斌

浙江大学宁波理工学院，宁波 (315100)

**摘要**：针对传统数据挖掘的不足，我们认为应当面向大数据开采"智慧"，并提出普适知识发现（UKD）新颖任务。与传统挖掘所获取的知识不同，普适知识具有一定的普遍性和不变性，易于形成洞察。本文首先定义了 UKD 的概念与内涵，指出它与科学定律之间的异同；然后，我们对普适知识进行了分类描述和刻画，比较了该类知识与传统挖掘获取的知识的特征差异；在此基础上，探讨了 UKD 的技术脉络，认为应当整合大数据挖掘、复杂性系统科学等相关领域的研究范式和技术方法，构建面向UKD的统一研究范式，并提出了 iBEST@SEE 方法论及其工作流程，从而为 UKD 的技术发展奠定了理论基础。

**关键词**：普适知识挖掘，智慧挖掘，范式转换，iBEST@SEE 方法论

**中图分类号**：TP 39    **文献标识码**： A    **文章编号**：

## 1. 引言

大数据（Big Data）时代，数据挖掘技术已经成为知识获取的关键手段。以"从数据中发现知识"为目标，遵循"数据预处理→数据抽样→模式挖掘→知识形成、后处理与知识理解"这一经典的知识发现流程[1]，近十年来，数据挖掘已经在技术上取得重大进展[2]，比如模式挖掘、聚类和分类、轨迹分析等。它们在对文本、WWW 数据、图、流数据等各种类型的数据处理上形成优势，并广泛应用于商务、社会管理、军事、人文、考古等各个领域。

然而，在应用数据挖掘所获取的知识时已经发现了如下问题：1）所得到的知识数量庞大，可理解性和可领悟性不足，在实际中往往无法有效地选择应用。以关联规则挖掘为例，动辄获得上百、上千条规则，当用户面对这么多的"规则"，其实很难从本质上对这些知识进行理解和消化。2）不同或者同一挖掘算法得到的知识质量参差不齐，知识之间存在不一致、甚至冲突，人们无法对知识的品质进行有效甄别[3]。3）对知识的普适性检验上存在欠缺，从而导致其可靠性受到质疑。例如，假设我们针对某一过去时间段的超市销售历史记录挖掘得到了经典的"购买'啤酒'=>购买'尿布'"规则，我们无法预知该规则在未来会不会突然失效，什么情况下会失效，规则有效性是如何波动的[4]？该规则对于其他超市是否适用？4）与领域知识结合不足，导致挖掘结果与实际应用需求存在鸿沟[5]。

我们认为突破上述局限的关键在于——数据挖掘技术需要把挖掘的目标更多地聚焦于普适知识和规律的发现上。人类对于信息获取和知识发现的目标与手段，已经经历了通过信

---

[*] 普适知识挖掘（Universal Knowledge Discovery, UKD）也可以称为普适信息挖掘，对应的英文可以为：Pervasive Knowledge Discovery （PKD），Pervasive Information Mining （PIM），Ubiquitous knowledge discovery 或者 Ubiquitous information mining 等。

息采集和检索技术获取信息，到使用数据挖掘生产普通知识两个阶段；下一个阶段，将面向"小数据"发现局域的、临时的、琐碎的粗糙知识上升到面向"大数据"挖掘全局的、长久的、精粹的普适知识，可以将数据挖掘理论与应用带入到一个新的阶段，它可以称为是"智慧发现"（Wisdom Discovery）。

因此，围绕"智慧发现"这一愿景，本文提出了一类新颖的知识发现任务：普适知识发现（Universal Knowledge Discovery，UKD），探讨了相关概念、技术脉络，并提出了 UKD 研究的方法论及其工作机制。

## 2. 什么是普适知识挖掘

对于数据挖掘而言，普适知识挖掘是一类新颖的知识发现任务。不同于传统数据挖掘方法得到知识的临时性和易变性，普适知识在大数据统计意义上，具有一定的普遍性和不变性，其应用优势在于：富有洞察性、可重复性和可预测性。为了清晰直观地介绍这一方向的新探索，我们在这里给出几个示例，来进行阐释和说明。

**例 1**. 商务：病毒式营销战略的监测和预测

营销经理希望在社交网络中设定若干个监测人员，通过监测人员所观察到的邻居局部信息，来监测一个病毒式营销在整个社交网络中的传播状况，预测它是否会流行。T. Chesney[6]通过重复多次的演化仿真，发现社交网络传播中的一些普遍性知识：传播内容在社交网络中的流行会呈现一条"S"形采纳曲线，是否超越一个采纳比例的临界值将成为流行与否的分水岭；一个监测节点的预测准确程度与它距离临界群体（即开始达到临界值时所有的采纳个体）中个体的最短距离负相关，与它的节点度中心性正相关等。根据这些发现，营销经理可以采取相应的行动，比如，选取中心性比较高的、分散在不同群（Cluster）中的节点作为观测点等。

**例 2**. 语言和脑科学：词汇普适关联分析及应用

经过各个语种本文大数据的分析，可以发现一系列的词汇间的普适的频繁关联模式，比如，"父亲"和"太阳"，在各个语言当中频繁一起出现。根据 RF i. Cancho[7]在《英国皇家学会会刊》上的研究，将频繁关联词汇彼此间建立连接，可以构成一个人类语言的小世界词汇网络；大脑通过神经元和突触联结来存储词汇及其网络[8]。因此，可以推测"父亲"和"太阳"概念在神经元连接上具有普适性。探究类似词汇普适性连接的根源，将有助于语言和脑科学的深入研究。

**例 3**. 跨学科：小世界和幂律分布规律

S. Milgram[9]，D.J. Watts[10]等分别从连锁信实验、演员合作网、电力网等大量实证数据中发现并验证了小世界现象，这种现象普遍存在在社会网络、Internet 网络、疾病传播等各个领域中；A.L. Barabási[11]在无标度网络中发现了节点度分布符合幂律分布的特点，而这种幂律分布规律的普适性已经被证实广泛存在于单词词频（Zipf 定律）、社会财富分布（Pareto 定律）、地震规模分布（Gutenberg - Richter 定律）、太阳耀斑强度的分布等社会和自然的各个方面，显示这一规律符合宇宙系统的演化规律。

归纳上述示例，我们给出具体的定义，描述如下。

**定义 1**. **（普适知识挖掘）**

普适知识挖掘是指从大数据中发现在统计意义上具有一定普遍性和不变性的知识与规

律。其中，普遍性是指知识所描述的现象比较普遍，不变性是指在一定时空、尺度等演化条件下不变。

需要加以说明的是：大数据统计意义上的普适性知识，包括但不限于科学定律（如图 1 所示）。普适知识在大数据统计意义上成立，它可以不像科学定律那么精确，但又非常有意义。比如，通过大规模手机数据发现，在灾难后，受灾人群的通讯行为往往发生改变：通讯量剧增，消息传递方式由普通情况下信源扩散式外传，变成事件亲历者与他们的重要联系人（例如，亲朋好友）之间频繁长时间双向通讯[12]。这一普适知识的发现，对于非常规突发事件应急管理具有实际意义。

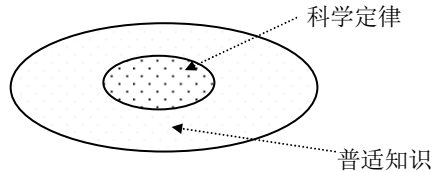

图 1. 科学定律与普适知识之间的差异

# 3. 普适知识的分类和刻画

### 3.1 普适知识的分类描述

首先，对所要挖掘的普适知识的基本类型进行梳理和归纳，它可能会随着大数据的获取和人类认知的深入而不断拓展，目前已知的、比较有代表性的、对人类获得洞察有帮助的普适知识类型有，但不限于以下方面：

**分布普适知识**。比较常见的数据分布有：正态/高斯、对数正态、指数/负指数、泊松、幂律、截断幂律、指数截断的幂律、漂移幂律、广延指数分布、Weibull 分布、胖尾分布等。比如，鱼类中的掠食者在食物富集时运动轨迹呈布朗运动，稀疏时采用 Lévy 飞行[13]等。

**性质普适知识**。比较常见的有：分形自相似性质、小世界、无标度、阵发性，例如 A.L. Barabási[14]认为大量的人类日常行为模式，从人们对维基百科的编辑，到货币经济公司的交易，都具有"阵发性"；金融市场上的长期记忆性（long-term memory）等。

**指数普适知识**。人们发现某些指标与现象存在着一定的关联，但不一定是因果关系，因此可以用指数对某些现象进行表征性描述，利用的是大数据的外部性。例如，"克强指数"用耗电量、铁路运货量和银行贷款发放量相结合，来评估经济状况。

**模式普适知识**。例如，①频繁共现模式；②顺序共现模式，例如，英文单词中，"I"和"am"频繁按顺序连续出现；③频繁序列模式，一个或多个事件在另外一个事件之后频繁发生；④行为模式[15]，比如行为的频次、时间间隔、互动对象、行为内容、行为时间等各个行为属性所呈现出的模式。

**规则普适知识**。可以是针对大数据获取到的普适性的关联规则、IF-THEN 规则、干预规则等。例如，通过分析 A 地区出生监测数据，发现叶酸充足的产妇所产的新生儿有先天缺陷的比例，远低于叶酸不足的产妇，从而得到一条干预规则[16]：

"Loc(A)∧Time(1995-2000) | 叶酸(不足→足) => 缺陷(有→无)，

支持度=0.33，变化度=0.33，置信度=0.5 "。

那么，如果进一步把这条规则在全国进行适用性检查，则可以得到普适干预规则，比如

是："叶酸(不足→足)=>缺陷(有→无)，普适性=全国，变化度=0.3±0.05"。

**关系普适知识**。不仅仅是因果关系，还包括更为广泛的正（负）相关关系、共现关系，它们可能属于同因不同果或同果不同因，或者仅仅是共同出现而已。而发现这些关系也是有意义的。例如，利用"流感"相关词条的检索与季节性流感爆发的相关性，Google Flu Trends可以更加及时、准确地预测流感的发生[17]。

**普适模型**。用机器模型、数学公式和物理模型等对普遍性现象进行归纳学习或者概括性描述，就可以得到普适模型。比如，"谷歌大脑"基于无监督的深度学习[18]，学会了区分和识别猫，这个习得的"神经网络"具有普适性，可以识别任何的猫。

**普适机理**。发现大数据所呈现普遍现象背后的普适机理，对人类获得洞察力具有显著意义。比如，A. L. Barabási[11]解释了增长和择优机制在复杂网络自组织演化中具有普遍性，它们使网络在宏观上具有幂律度分布的普适现象。

### 3.2 普适知识和传统数据挖掘获取的知识的比较

基于前述对普适知识的分类描述和刻画，下面，我们对两种知识进行特征比对，如表 1 所示。可以发现，与传统挖掘所获取的知识相比，普适知识具有显著不同的特点：①范围更广，类型更为多样；②从系统、整体和宏观层面提取知识；③精确和粗略描述均可；④面向大数据实施发掘、检验；⑤易于获得洞察，具有较强的可行动性。

表 1. 两种知识的特征对比

| 比较项 | 传统数据挖掘获取的知识 | 普适知识 |
| --- | --- | --- |
| 类型 | 模式、规则、机器模型等 | 可以是分布、性质、指数、关系等等，也涵括模式、规则和模型。 |
| 描述层次 | 一般是微观、中观 | 系统、宏观、整体 |
| 精准程度 | 大多数精确、细致 | 精确描述可以，大致粗略的描述亦可 |
| 数据范围 | 通常小数据、局部数据 | 大数据，系统全局的、整体的数据 |
| 检验方式 | 抽取部分样本作检验和评价 | 大数据检验 |
| 通用性 | 绝大多数未经进一步通用性检验 | 具有一定程度的普适性 |
| 可理解性 | 偏弱 | 强（容易获得洞察） |
| 可行动性 | 偏弱 | 强（易于采取行动） |
| 计算方式 | 往往是精确计算 | 估算亦有效 |

## 4. 普适知识挖掘的技术渊源和范式探讨

根据前述普适知识的描述与刻画，接下来的问题是：如何面向大数据挖掘普适知识。我们认为普适知识挖掘需要充分吸收和整合大数据挖掘、复杂性系统科学等相关学科的前沿进展，并形成适合于普适知识挖掘的新型研究范式。

### 4.1 从"机器发现"到"数据挖掘"，以及大数据条件下数据挖掘的研究动态

早在上世纪 70 年代末，P. Langley[19,20]就开始尝试采用启发式搜索的方法，相继研发了 6 个版本的 BACON 系统，进行某些物理和化学定律的自动再发现（rediscovery）；此后，借助

于归纳学习和逻辑推理等方法，又相继诞生了 STAHL[21]、FAHRENHEIT[22]、IDS[23]等系统，我国学者则在几何定理的机器证明[24]等方面取得了丰硕的成果，从而在上世纪 80 年代形成了"机器发现（Machine discovery）"的研究方向。其局限在于：1）要有比较纯净的少量数据作为输入和一定的搜索定向，否则过大的搜索空间将导致算法不可行；2）仅仅再发现了已知定律，而没有（或很少）发现新的、重要科学定律。

自上世纪 90 年代中叶起，数据挖掘整合统计和机器学习的方法，避开机械发现的局限，着力发现一些有意义的模式知识（Pattern），并得到广泛应用。从而，掀起 KDD 研究热潮。

随着大数据时代的到来，当前数据挖掘和大数据解析（Big data analytics）领域关注于处理大数据的海量性（Volume）[25, 26]、高速增长（Velocity）、数据类型的多样化（Variety）等方面的技术[27]，重点探讨和比较所提出方法的性能优劣、可扩展性和适用范围，而对大数据条件下挖掘目标的升华、大数据挖掘在思维和方向上的根本性转换[28]，以及由此而来的挖掘范式、方法和手段的根本性变革尚缺乏足够的、深入的认知和思考。

比如，著名咨询机构麦肯锡公司[29]在报告中清晰地指出，"大数据：下一个创新、竞争和生产力的前沿地带"，但在所采用的技术上，仍然沿袭传统数据挖掘的方法；J. Cohen[30]等提出了 MAD（Magnetic 吸引性、Agile 敏捷的、Deep 深入的）大数据分析技巧，用于替代传统的商业数据仓库分析方法；P. Zikopoulos[31]着重介绍了用于大数据分析的流行的 Hadoop 分布式处理平台。2013 年 SIGKDD 知识发现会议上"Big data framework"主题的三篇文章均围绕大数据的高性能处理，分别提出了一种单一 PC 上处理十亿规模的并行图形引擎[32]、一种大数据管道（Pipelines）的概率框架，和一个集成软硬件和设计模式的 BID 数据套件[33]。W. Fan[34]在综述大数据挖掘现状和趋势时，着眼于大数据的计算可处理性，以及处理不同类型的大数据，如挖掘异构信息网络（Mining Heterogeneous Information Networks）等。

虽然大数据的可处理性是大数据管理和挖掘中的重要方面，但是不是这样就能从大数据中萃取出"黄金"，获得大洞察？显然，与小数据挖掘相比，大数据挖掘不仅仅是能处理的数据量的增长，更重要的是体现在数据挖掘思维和技术范式的转换，使其具备举重若轻、行云流水般的数据处理能力，既能像处理小数据那样处理大数据，又能采用曲径通幽的方式，通过数据的外部性巧妙地获得洞察和智慧[35]。我们认为下面这些工作具有典型意义，对于大数据挖掘比较具有启发性：

例如，2014 年春节，央视利用 2 亿多部智能手机移动大数据，用可视化的方式展示了春运中人流迁移的宏观统计规律，既易于相关部门合理地调度春运资源，又便于个人更好地安排出行计划[36]。这里，充分利用了智能手机轨迹数据的外部性。Y.-X. Zhu[37]等分析了中华菜谱大数据，认为决定菜系形成的主要因素是地理上的邻接和相近，而不是气候差异。这里，所分析的数据是全国菜系的数据，是全体数据，而不是抽样数据，当然如果分析的是全世界的菜谱大数据，所得出的结论会更有说服力。F. Zhang[38]等研究了通过 GPS 感知城市出租车位置的移动情况，探测出租车加油站加油所花费的时间，从而为出租车选择加油站提供推荐，同时以此来估算整个城市实时的耗油量。这里使用了智慧城市条件下整个城市的泛在感知动态性数据，包含了实时调度和决策的特征。

这些大数据实践工作相当出色，但是，总体而言，对于大数据的挖掘应用尚处于个例摸索的阶段，还没有形成一种理想的大数据挖掘技术范式。

## 4.2 基于大数据的复杂性系统科学实践进展与范式特点

在国内，早在上世纪 80 年代初，钱学森[39, 40]、许国志[41]等就开始倡导系统科学，但是由于数据的匮乏，很多重要思想和观点得不到实证数据的有力支持。近年来，随着大数据采集、存储和分析技术的发展，使得面向大规模数据探寻机制和规律，以及在大数据上进行统计验证成为可能，从而推动复杂性系统科学进入到一个新的高潮阶段。

比如，D.J. Watts[42]发表在 Nature 上的小世界模型和 A.L. Barabási[11]在 Science 上的无标度网络这两个标志性工作的诞生，均受益于演员合作网、WWW 网等大规模网络实证数据的获得。复杂网络的反问题，也就是通过动力学表征来反推网络结构，像链路预测[43]等，所提出的方法都需要经过实际大型网络的验证。再例如，D. Brockmann[44]通过分析 46 万张带标记的一美元钞票的流通轨迹，来反映较大空间尺度上人类的移动特征，发现钞票移动步长在一定范围内符合标度律。此后，M.C. González[45]、阎小勇[46]、C.-B. Peng[47]等分别采用 10 万手机用户移动轨迹、230 个志愿者 6 星期的旅行日志、上海 158 万出租车运营路径来分析人类的出行行为，并得出许多有益的结论。

可以发现，与数据挖掘方法相比，上述复杂性系统科学研究完全采用了另外一种技术进路，其范式特点归纳如下[48]：

1）系统论的观点和网络的视角

摒弃还原论，采用系统论的范式，提供了一套网络结构的视角，对系统进行抽象化的建模描述。在此基础上，探讨动力学、功能涌现等核心问题。它往往从微观的简单动力学规则入手，再现和剖析宏观现象的机理。

2）演化的观点与动力学探讨

从演化的观点来看待复杂系统，既讨论系统结构演化动力学，也包含了系统所承载的动力学。以复杂网络为例，在探讨静态网络结构基础上，探索丰富多彩的动力学机制，比如：传染病传播、舆情传播、级联失效、网络拥塞、人肉搜索等。

3）大量使用统计物理的方法

一方面，统计物理发挥了重要的功效，提供了大量的有效方法，比如：平均场理论、渗流模型、主方程、Fokker-Planck 方程等等。另一方面，复杂性研究也关注系统中发生的物理过程和物理性质，比如：玻色–爱因斯坦凝聚、随机游走等。

4）仿真与大数据统计相结合

面向大数据探寻、发掘和验证统计规律，剖析内在机理；同时，通过计算机仿真的形式进行推演，探测复杂系统宏观特性的涌现。

综上，复杂性系统研究范式在系统建模、动力学演化、内在规律和机理的探索等方面的优势是显而易见的，但是在分析和处理大数据方面存在着瓶颈。汪秉宏等[49]认为"有组织、有针对性地从实验实证数据探索系统的内外约束条件还处于初步阶段"，"关键的问题是缺乏有效的理论框架，所获得的数据在数量上可能很大，却不一定有很强的针对性"。实际上，数据挖掘技术在数据采集和处理，以及数据驱动的模型研究上，有许多成功的方法和经验可供借鉴[48]。

## 4.3 大数据挖掘和系统科学研究范式的比较

根据前述讨论，我们可以把大数据挖掘和系统科学这两个领域中的研究范式做一个比较，如表 2 所示。概括地讲，复杂性系统科学研究从整体的、演化的视角出发，采用内在机理分析的方式，擅长于发现宏观规律；大数据挖掘研究采用数据驱动的方式建立模型，擅长于处理各类数据，与领域实际应用结合紧密。因此可以发现，在大数据时代，两个领域的研究实际上具有很强的互补性。它们殊途同归，共同指向大数据分析和挖掘的下一站——普适知识挖掘。

表 2. 两个领域中的研究范式对比[48]

| 统称 | 大数据挖掘 | 系统科学 |
| --- | --- | --- |
| 关注点 | 大数据的可处理性、平台和算法的性能、所发现知识的新颖性等 | 系统结构、动力学与功能三要素 |
| 视角 | 基于大数据的归纳、统计和机器学习 | 系统的、网络连接的、演化的视角和观点 |
| 思维 | 全体或抽样数据，利用外部性、估算或精确计算、寻找相关关系或因果关系 | 复杂性思维，如自组织、涌现、自相似、混沌等 |
| 模型 | 机器学习模型、部分数学模型 | 数学、物理模型 |
| 主要方法 | 统计分析，挖掘算法、机器学习 | 统计物理、动力学方程、仿真与部分算法 |
| 知识类型 | 机器习得模型、模式（Pattern）、规则 | 数学物理模型、不同层次上的普适规律、内在机制和机理 |
| 长处 | 擅长处理各类静态和动态的大数据：文本、图像、流数据等，以及异质数据，易于习得或发现隐藏的模式 | 擅于发现模型、规律，易于获得理解和认知 |

# 5. iBEST@SEE 方法论：普适知识挖掘的跨学科范式整合

大数据时代，面对普适知识发现这一类新颖的数据挖掘任务，迫切需要对各个学科和领域所发展出来的研究范式和技术方法做一个跨学科的有机整合，特别是要充分吸收系统科学和大数据挖掘的研究进展，以利于多学科知识和方法的交叉和融合。因此，我们提出了 iBEST@SEE 方法论及其工作流程，用于实现普适知识挖掘研究范式的跨学科整合。

## 5.1 iBEST@SEE 方法论的内涵

首先，我们分别用大数据挖掘（Big Data Mining）（含数据驱动的分析、数理统计等）和系统科学（System Science）（含复杂性系统思维、系统建模、系统动力学解析等）来概括上述两个领域中的研究范式。此外，我们还需要从系统科学方法体系中专门抽取出"仿真模拟"研究方法，同时借鉴和吸收其他学科和领域中一些比较经典的普适知识探索方法（或范式），描述如下：

1）仿真模拟（Simulation）

使用计算机或其它仿真系统模拟真实系统进行实验研究，尤其适合多主体仿真、系统动力学演化、数值模拟等，具有快速、直观和成本低廉的优势。

2）可控实验（Experiment）

在环境不变（实验室或部分真实环境）的条件下，控制一个或多个变量，得到因变量相

应变化以检验假设。它包含了前提（含假设）、实验操作、结果验证等步骤。

3）理论分析（Theory Analysis）

充分利用领域理论，运用数学、物理等工具，进行理论建模、解析、逻辑演绎、公式推演和证明，用于得出推论，理解模型，仿真和实验的假设、过程和结果等。

4）人机智能（Intelligence）

人机智能包括了三部分方法：一是机器智能（含专家系统、机器学习和 Web 智能等），二是人类智能（含个体智能、集体智能等），三是人机交互与协同（含可视化）。

5）实证研究（Empirical Research）

基于大数据的普适知识发现属于一个典型的实证问题。因此，需要采用实证循环[50]的方式进行检验："观察（搜集和组织经验事实）→ 归纳（提出假设）→ 建模（构建模型）→ 演绎（形成可测试推论）→ 测试（验证推论）→ 评估（评估假设和模型）"，然后进入下一轮循环，逐步调整假设与模型，使其更好地吻合事实。

基于前述讨论，我们提出了挖掘普适知识的 iBEST@SEE 方法论。"iBEST@SEE"是基于"人机智能（intelligence）+ 大数据挖掘（Big Data Mining）+ 可控实验（Experiment）+ 仿真模拟（Simulation）+ 理论分析（Theory Analysis）@ 系统科学（System Science）+ 扩展的实证研究框架（Extended Empirical Research Framework）"而形成的。这里"@"强调从系统科学的视角出发，使用实证研究的逻辑主线有机组织各种方法；"i"采用小写以示区分，用于强调人类智慧在整个创造性发现中的作用。

## 5.2 普适知识挖掘的工作范式

由于传统的知识发现流程[1]已很难适应普适知识挖掘的需要。因此，基于 iBEST@SEE 方法论，我们提出了普适知识挖掘的三种工作范式，阐述如下：

**范式 1**. 大数据"单环"驱动的挖掘模式

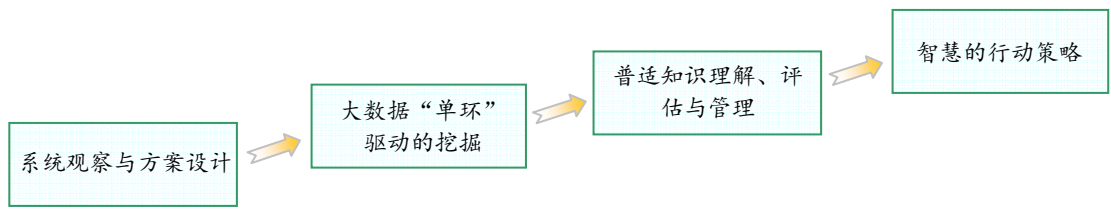

图 2. 大数据"单环"驱动的普适知识挖掘流程

在已经具备相应的大数据以及明确的普适知识挖掘方向时，可以直接采用大数据解析"单环"驱动的挖掘模式。它包括"系统观察与方案设计→大数据'单环'驱动的挖掘→普适知识理解、评估度量与管理→智慧的行动策略"等四个环节（如图 2 所示）。

这里，"系统观察"是指确定要研究的系统对象，观察系统所呈现的宏观现象与功能，提出要挖掘和探索的领域目标；"方案设计"是指对整个普适知识挖掘过程的设计与安排，因为大数据挖掘需要具有思维上的转换与创新，此外，还包括目标导向的大数据积累、准备等方面的整体设计；"大数据'单环'驱动的挖掘"是核心成分（如图 5 的左环所示），按照"大数据采集与记录→大数据预处理→大数据分析和挖掘→普适现象萃取→现实预测、验证与评估"五个步骤构成螺旋式提升的过程，从大数据中萃取得到具有普适性的统计规律；

"普适知识理解、评估与管理"是后续普适知识利用工作的基点，它对所得到的知识进行可用性（适用范围和可靠性）评估度量，并且作为战略资产进行有效管理（专有、分享和传播等）；最后，"智慧的行动策略"指围绕领域问题，依据所发现的规律，结合领域知识，生成智慧行动的策略。

大数据"单环"驱动的挖掘模式比较适合于具有清晰定向的学习、搜索和验证式普适知识发现，比如：挖掘分布、性质、模式、规则、机器模型等类型的普适知识。

**范式2．机制"单环"驱动的分析模式**

在缺乏实证大数据的情况下，可以使用"机制'单环'驱动的分析模式"探索性分析实验、仿真等手段所获取的实验数据。它包含"系统观察与方案设计→机制'单环'驱动的挖掘→规律提出以及进一步的大数据检验→普适知识的管理与利用"等四个环节（图3）。

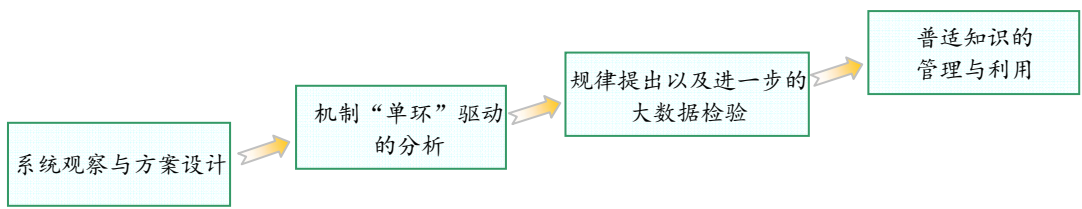

图 3. 机制"单环"驱动的普适知识发现流程

其中，第一步"系统观察与方案设计"与流程 I 中所述相同；第二步"机制'单环'驱动的挖掘"是核心（如图 5 右环所示），采用实证的逻辑框架，按照"系统抽象和建模→理论分析和假设、机理提出→仿真模拟和受控实验→分析实验数据→涌现或发现普适现象"五个环节构成螺旋式渐进的过程。如果假设和机理能够很好地解释实验中观测到的现象，则可接受它们，否则进一步修正，直至可以很好解释为止；而实验中的现象也要能够与现实观测相吻合。这样，进入到第三步，"规律提出以及进一步的大数据检验"，这里的规律包括"机制环"所发现的普适现象和提出的假设、机理，同时进一步接受实证数据的检验，直至得到公认。最后，"普适知识的管理与利用"就是范式1中的后面两个环节。

机制'单环'驱动的分析模式适合于探索性地发现各类普适知识，比如数学和物理模型等。

**范式3．大数据和机制"双环"协同驱动的挖掘模式**

"双环"驱动的挖掘过程（图 4）与大数据"单环"驱动的流程基本相同，不同之处在于使用"大数据和机制'双环'运作的模式"替代了"单环"的工作方式。它是整个流程中最为核心的部分，如图5所示。通过"双环"协同驱动，发现大数据普适现象，同时挖掘既能解释现象，又能预测现象的深层次机制和规律。

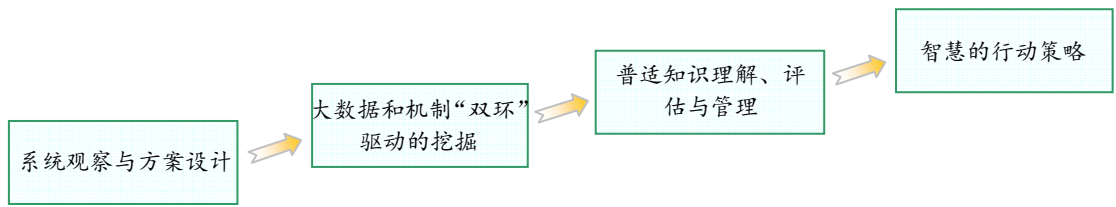

图 4. 大数据和机制"双环"协同驱动的挖掘流程

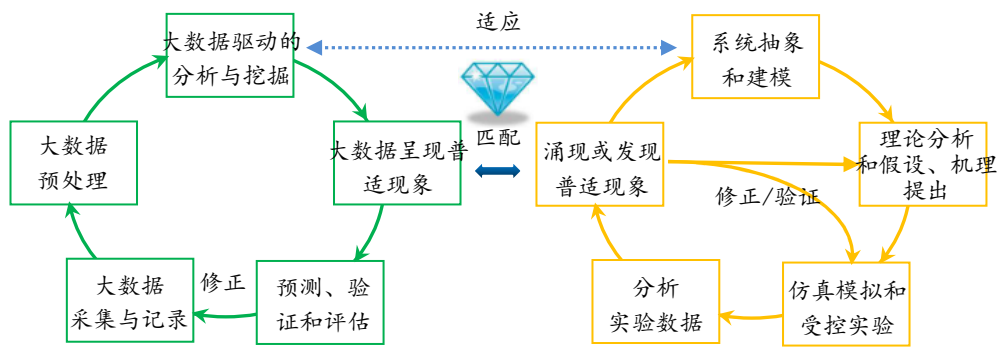

图 5. 大数据和机制"双环"协同驱动的普适知识挖掘模式

这里的"双环"协同体现在两个协同、两个匹配上。两个协同：① 机制环的"系统抽象和建模"指导数据环的"大数据采集、记录、分析和挖掘"，数据环的"数据采集、记录和挖掘"建构在机制环的"系统建模"之上，且反映"系统建模"；② 数据环的"大数据呈现普适现象"支持机制环"假设、机理提出"，机制环"假设、机理提出"引导"大数据中发现规律"。两个匹配：① 数据环中"大数据呈现的普适现象"要与机制环中通过"仿真和实验"再现的"普适现象"相匹配；② 数据环中记录和观察到的"宏观现象"要与机制环中通过"仿真和实验"涌现和发现的现象相匹配。

该挖掘模式的优势在于通过"双环"协同、相互启发，螺旋式渐进地探索大数据层面的普适规律，以及系统内部的工作机理。它适合于挖掘各类普适知识及其背后的机理。

## 6. 总结和讨论

围绕"智慧发现"这一愿景，本文提出了一类新颖的知识发现任务——普适知识挖掘（Universal Knowledge Mining，UKM）。首先，定义了普适知识挖掘的概念与内涵，进而对普适知识进行了分类描述和刻画，阐述了该类知识的显著特征；在此基础上，探讨了普适知识挖掘的技术脉络，提出应当整合大数据挖掘、复杂性系统科学等相关领域的研究范式和技术方法，构建统一的面向普适知识挖掘的 iBEST@SEE 方法论；提出了 iBEST@SEE 方法论的内涵，并阐述了 iBEST@SEE 三种工作流程，从而为未来 UKM 的技术研发奠定了坚实的方法论基础。

未来进一步的工作包括：如何围绕各类不同的普适知识，提出相应的挖掘技术和方法，以及如何利用大数据和机制"双环"协同驱动机制，深度挖掘"智慧"等。可以预见，从大数据中，人类会发现越来越多的普适知识。它们将会极大地改变人类的生产和生活方式，从而使人类进入"智慧"时代。

## 参考文献

[1] U. Fayyad, G. Piatetsky-Shapiro, P. Smyth. From data mining to knowledge discovery: an overview[J]. *AI Magazine*, 1996, 17: 37-54.

[2] J. Han, M. Kamber, J. Pei. *Data mining: concepts and techniques* [M]. 3rd edition. Burlington: Morgan Kaufmann, 2011.

[3] 李兴森, 石勇, 张玲玲. 从信息爆炸到智能知识管理[M]. 北京：科学出版社, 2010.

[4] B. Shen, M. Yao, Z. Wu, Y. Gao. Mining Dynamic Association Rules with Comments[J]. *Knowledge*

and *Information Systems*, 2010, 23(1): 73-98.

[5] L. Cao, P. S. Yu, C. Zhang, Y. Zhao. *Domain driven data mining*[M]. Berlin: Springer, 2010a.

[6] T. Chesney. Networked individuals predict a community wide outcome from their local information[J]. *Decision Support Systems*, 2014, 57: 11-21.

[7] RF i. Cancho, R.V. Sole. The small world of human language[J]. *Proceedings of the Royal Society B (Biological Science)*, 2001, 268(1482): 2261-2265.

[8] F. Pulvermüller. *The neuroscience of language: on brain circuits of words and serial order*[M]. Cambridge: Cambridge University Press, 2003.

[9] S. Milgram. The small-world problem[J]. *Psychology Today*, 1967, 1(1): 61-67.

[10] D. J. Watts, S. H. Strogatz. Collective dynamics of 'small-world' networks[J]. *Nature*, 1998, 393: 440-442.

[11] A. L. Barabási, R. Albert. Emergence of scaling in random networks[J]. *Science*, 1999, 286(5439): 509-512.

[12] L. Gao, C. Song, Z. Gao, A.L. Barabási, J.P. Bagrow, D. Wang. Quantifying information flow during emergencies[J]. *Scientific Reports*, 2014, 4: 3997(1-6).

[13] G.M. Viswanathan. Fish in Lévy-flight foraging[J]. *Nature*, 2010, 465: 1018-1019.

[14] A.L. Barabási. Bursts: *The hidden patterns behind everything we do, from your e-mail to bloody crusades*[M]. New York: Penguin Group, 2011.

[15] L. Cao, P.S Yu. *Behavior computing: modeling, analysis, mining and decision*[M]. Berlin: Springer, 2012.

[16] 唐常杰, 张悦, 等. 亚复杂系统中动力学干预规则挖掘技术研究进展[J]. *计算机应用*, 2008, 28(11): 2732-2736.

[17] J. Ginsberg, M.H. Mohebbi, R.S. Patel, L. Brammer, M.S. Smolinski, L. Brilliant. Detecting influenza epidemics using search engine query data[J]. *Nature*, 2009, 457: 1012-1014.

[18] Y. Bengio. Learning deep architectures for AI [J]. *Foundations and trends in machine learning*, 2009, 2(1): 1-127.

[19] P. Langley. BACON.1: A general discovery system [C]// *Proceedings of the Second Biennial Conference of the Canadian Society for Computational Studies of Intelligence*. Toronto, 1978.

[20] P. Langley. Rediscovering physics with BACON.3 [C]// *Proceedings of the Sixth International Joint Conference on Artificial Intelligence*. Tokyo, Japan, 1979.

[21] J.M. Zytkow, H.A. Simon. A theory of historical discovery: the construction of componential model [J]. *Machine Learning*, 1986, 1(1): 107-136.

[22] J.M. Zytkow, J. Zhu, A. Hussam. Automated discovery in a chemistry laboratory [C]// *Proceedings of AAAI'90*, Boston, Massachusetts, 1990.

[23] B. Nordhausen, P. Langley. An integrated framework for empirical discovery [J]. *Machine Learning*, 1993, 12(1-3): 17-47.

[24] 吴文俊. 几何定理机器证明的基本原理[M]. 北京: 科学出版社, 1984.

[25] J. Lin, D. Ryaboy. Scaling big data mining infrastructure: the twitter experience[J]. *SIGKDD Explorations*, 2012, 14(2): 6-19.

[26] A. Rajaraman, J. D. Ullman. *Mining of massive datasets*[M]. Cambridge: Cambridge University Press, 2012.

[27] IBM. What is big data ?[EB/OL]. [2012-12-02]. http://www-01.ibm.com/software/data/bigdata/.

[28] Viktor Mayer-Schönberger, Kenneth Cukier. *Big data: a revolution that will transform how we live, work, and think*[M]. Eamon Dolan / Houghton Mifflin Harcourt, 2013.

[29] McKinsey Global Institute. *Big data: The next frontier for innovation, competition, and productivity*[R]. [2011-3-20].

http://www.mckinsey.com/insights/business_technology/big_data_the_next_frontier_for_innovation.

[30] J. Cohen, B. Dolan, M. Dunlap, J. M. Hellerstein, G. Welton. MAD skills: new analysis practices for big data[C]// *Proceedings of VLDB'09*, Lyon, France, 2009.

[31] P. Zikopoulos, C. Eaton. *Understanding big data: analytics for enterprise class hadoop and streaming data*[M]. New York: McGraw-Hill Osborne Media Publisher, 2011.

[32] W.-S. Han, S. Lee, K. Park. Kim, et al. TurboGraph: a fast parallel graph engine handling billion-scale graphs in a single PC[C]// *Proceedings of KDD'13*, Illinois, USA, 2013.

[33] J. Canny, H. Zhao. Big data analytics with small footprint: squaring the cloud[C]// *Proceedings of KDD'13*, Illinois, USA, 2013.

[34] W. Fan, A. Bifet. Mining big data: current status, and forecast to the future[J]. *SIGKDD Explorations*, 2013, 14(2): 1-5.

[35] 周涛. 大数据时代，中国并未落后[EB/OL]. 科学网, [2013-01-29].
http://blog.sciencenet.cn/blog-3075-657481.html.

[36] 唐常杰. 假日聚会, 戏说云物人海[EB/OL]. 科学网, [2014-02-08].
http://blog.sciencenet.cn/home.php?mod=space&uid=287179&do=blog&id=765603

[37] Y.-X. Zhu, J.-M. Huang, Z.-K. Zhang, et al. Geography and similarity of regional cuisines in China[EB/OL]. [2013-7-11]. http://arxiv.org/pdf/1307.3185v1.pdf.

[38] F. Zhang, D. Wilkie, Y. Zheng, X. Xie. Sensing the pulse of urban refueling behavior[C]// *Proceedings of UbiComp'13*, Zurich, Switzerland, 2013.

[39] 钱学森. 再谈系统科学的体系[J]. *系统工程理论与实践*, 1981, 1(1): 1-3.

[40] 钱学森, 于景元, 戴汝为. 一个科学的新领域：开放的复杂巨系统及其方法论[J]. *自然杂志*, 1990, 13(1): 3-10.

[41] 许国志, 顾基发, 车宏安. *系统科学*[M]. 上海: 上海科技教育出版社, 2000.

[42] D. J. Watts, S. H. Strogatz. Collective dynamics of 'small-world' networks[J]. *Nature*, 1998, 393: 440-442.

[43] 吕琳媛, 周涛. *链路预测*[M]. 北京: 高等教育出版社, 2013.

[44] D. Brockmann, L. Hufnagel, T. Geisel. The scaling laws of human travel[J]. *Nature*, 2006, 439(26): 462-465.

[45] M.C. González, C.A. Hidalgo, A.L. Barabási. Understanding individual human mobility patterns[J]. *Nature*, 2008, 453: 779-782.

[46] X.-Y. Yan, X.-P. Han, B.-H. Wang, T. Zhou. Diversity of individual mobility patterns and emergence of aggregated scaling laws[J]. *Scientific Reports*, 2013, 3: 2678.

[47] C.-B. Peng, X.-G. Jin, K.-C. Wong, et al. Collective human mobility pattern from taxi trips in urban area[J]. *PLOS One*, 2012, 7(4): e34487.

[48] 沈斌. 复杂网络与数据挖掘:研究范式的比较和整合[J]. *复杂系统与复杂性科学*, 2014, 11(1): 48-52.

[49] 汪秉宏, 周涛, 周昌松. 人类行为、复杂网络及信息挖掘的统计物理研究[J]. *上海理工大学学报*, 2012, 34(2): 103-117.

[50] A.D. de Groot. *Methodology: foundations of inference and research in the behavioral sciences*[M]. Belgium: Mouton & Co., 1969.